# Understanding and Mitigating Banking Trojans: From Zeus to Emotet

Konstantinos P. Grammatikakis*, Ioannis Koufos*, Nicholas Kolokotronis*, Costas Vassilakis*, Stavros Shiaeles†
*Department of Informatics and Telecommunications, University of the Peloponnese, 22131 Tripolis, Greece
kpgram@uop.gr, ikoufos@uop.gr, nkolok@uop.gr, costas@uop.gr
†Cyber Security Research Group, University of Portsmouth, PO1 2UP, Portsmouth, UK
stavros.shiaeles@port.ac.uk

*Abstract*—Banking Trojans came a long way in the past decade, and the recent case of Emotet showed their enduring relevance. The evolution of the modern computing landscape can be traced through Emotet and Zeus, both representative examples from the end of the past decade. As an example of earlier malware, Zeus only needed to employ simple anti-analysis techniques to stay undetected; while the more recent Emotet had to constantly evolve to stay a step ahead. Current host-based antimalware solutions face an increasing number of obstacles to perform their function. A multi-layer approach to network security is necessary for network-based intrusion response systems to secure modern networks of heterogeneous devices. A system based on a combination of a graphical network security model and a game theoretic model of cyber attacks was tested on a testbed with Windows machines infected with Trojans; experimental results showed that the proposed system effectively blocked Trojans' network communications effectively preventing data leakage and yielding encouraging results for future work.

## I. Introduction

The international collaborative effort to take down Emotet's infrastructure in early 2021 [1] brought Trojans back to the spotlight. First detected in 2014, Emotet started as banking Trojan, but soon became one of the the most prevalent malware campaigns of the recent past [2], [3]. At the time of its infrastructure take-down, Emotet had evolved to a full crimeware service with a modular structure, having developed the ability to perform a number of actions, besides its core functionality as a banking Trojan, which included the installation of other malware (e.g. TrickBot, which itself was responsible for 31% of the global banking malware detected in 2019 [2]). In addition to its complex structure, Emotet employed multiple layers of countermeasures against analysis efforts and antimalware solutions [4], [5].

Almost a decade ago, in late 2010 [6], a similar operation was organized to take down another infamous banking Trojan. Starting in a landscape dominated by personal computers, in the latter half of the 2000s, Zeus reached infamy with 3.6 mil infections in the United States [7] in 2009. After the disclosure of its full source code in mid 2011, Zeus' development branched out, producing a number of successful and more sophisticated variants, including Citadel and GameoverZeus [8]. The original branches of Zeus throughout their lifetime used standard methods to capture credentials (namely, web injects and keystroke logging), basic countermeasures against detection attempts [9] and data-based obfuscation processes [7]. In late 2010, as two-factor authentication schemes were introduced by financial institutions and smartphones became increasingly popular, the Zeus crimeware kit was expanded with *Zeus-in-the-Mobile* (ZitMo), a companion malware targeting mobile devices to intercept *mobile transaction authentication numbers* (mTAN) received via *short message service* (SMS) messages [10].

The stories of Zeus and Emotet run parallel to the evolution of the computing landscape over the past decade, which shifted fundamentally from the homogeneous market of personal computers (i.e. desktops and laptops), to today's diverse market of smartphones, tablets and other smart appliances. According to the latest data published by the Pew Research Center in the United States [11], [12], ownership of smartphones and tablets among the surveyed population increased dramatically to 81% and 52% (from 35% and 8% in 2011), respectively, while Internet access has reached 90% (from 79% in 2011).

The use of host-based solutions (e.g. antimalware systems, or personal firewalls) to defend against malware has been the subject of intense research for decades, but their effectiveness varies between different kinds of computing devices [13]. As evidenced by incidents of the past decade, this heterogeneous landscape requires a multi-level approach to network security, by a combination of host-based and network-based solutions. Such network-based solutions fall under the category of *intrusion response systems* (IRS), which are able to detect and respond to network attacks by selecting the most appropriate mitigation actions, balancing the effectiveness of the chosen actions with service availability [14].

This paper presents the infection techniques, exhibited behavior and network communication patterns of two representative banking Trojans of the previous decade, Zeus (along with its companion ZitMo) and Emotet, so as to provide further insight on the future of banking Trojans and of malware in general. Additionally, to test the effectiveness of IRS solutions

This project has received funding from the European Union's Horizon 2020 research and innovation programme under grant agreement no. 786698. The work reflects only the authors' view and the Agency is not responsible for any use that may be made of the information it contains.

to defend against such malware, an IRS based on a *graphical network security model* (GNSM) of the network and a game theoretic model of cyber attacks was tested on the presence of a Zeus-infected device. The results of its testing were encouraging, as Zeus network traffic was successfully blocked, but further testing against newer threats, including Emotet, is required to verify its effectiveness against novel threats. The paper's main contributions are as follows:

- Insight on the rapid increase in malware complexity is provided, by comparing the attack strategies and employed countermeasures of two representative examples from the past decade.
- Common behavioral patterns between Zeus and Emotet were identified, including the use of encrypted network communications which, when implemented correctly, can bypass signature-based detection systems.
- Automatically generated intrusion response strategies have been successfully tested against Zeus-infected Windows machines in a realistic network testbed.

The rest of the paper is organized as follows: previous work on Zeus, ZitMo and Emotet is presented in Section II; a high-level presentation of their infection techniques, behavior and network communications is given in Section III; whilst more technical details are presented in Section IV. The IRS and the main findings of its testing are presented in Section V, while concluding remarks are provided in Section VI.

## II. Related work

Interest in the area of malicious software research has been steadily increasing for decades, and Zeus quickly became the subject of intense research in its own right, as the scale of its impact grew significantly and its targets became global. One of the earliest reports was presented by [15], covering the usage of the toolkit (ver. 1.2.5.4) and some observations on a (then) live campaign which used the Fiesta web exploitation framework. A more complete report was presented by [9], covering the usage and inner workings of the toolkit (ver. 1.2.4.2), its information gathering functionalities, additional capabilities and network communication patterns of the produced bots, along with a review of the *command and control* (C&C) server functionality. A reverse engineering effort on the same version of the toolkit was presented by [7], in which the four obfuscation layers were analyzed, a script to automate the encryption key recovery from a captured bot executable was created, and the possibility to utilize false information injection as an active countermeasure was discussed. A comparative study of two versions of the toolkit (ver. 1.2.7.19 and 2.0.6.5) was conducted by [16], noting the evolution of the toolkit that allowed a single machine to be infected multiple times by different botnets. A detailed analysis of the first publicly disclosed version (ver. 2.0.8.9) was presented by [17], a detailed presentation of its obfuscation algorithm, the usage of RC4 encryption and the key extraction process from captured network traffic were presented, while an *intrusion detection system* (IDS) specifically for Zeus was presented and benchmarked. GameoverZeus, a variant following a decentralized networking approach was described by [18], [19]; an analysis of samples gathered over a two-year period allowed for a detailed presentation of its peer-to-peer protocol and of the three-layer topology formed by its botnet. Authorship analysis on the first publicly disclosed version (ver. 2.0.8.9) was conduced by [20], indicating that the toolkit was developed by two persons, each working on modules across the entire source code tree. An investigation on the uptime of the C&C servers for Zeus, Citadel, Ice IX and KINS botnets, presented by [21], indicated that the geographic location and hosting type greatly affected uptime. A comparative review of six banking malware was conducted by [22] to track their use of anti-analysis measures; using a version 2 sample as a representative of early banking Trojans, the authors noted the increasing effort needed to reverse engineer such malware, and the choice of simple, yet effective, measures by knowledge of the malware analysis process; such processes typically involve static and dynamic analysis techniques [23].

Similar research interest was also shown for ZitMo; being one of the first malware targeting early mobile devices, it showed the future direction for malicious actors and security researchers alike. A detailed report about the first observed version of ZitMo was presented by [10] and updated in [24], the reports noted the novelty of mobile Trojans, ZitMo's narrow aim to only intercept mTAN messages, along with its wide range of targeted platforms. The results of reverse engineering efforts on a sample targeting Symbian devices was presented by [25]; the close relation of this version with SMS Monitor, an existing spyware, was shown by the large percentage of common strings and assembly routines. A later version of ZitMo for Android was analyzed by [26], [27], it was discovered that this version had the ability to communicate both via SMS messages and via *hypertext transfer protocol* (HTTP) requests. An evaluation of the effectiveness of eleven antimalware solutions and of the actual risk imposed by Android malware was presented by [13]; it showed that antimalware solutions are ineffective against unknown malware, which resulted in only one antimalware solution being able to detect altered ZitMo samples. An Android malware classification system was presented by [28], using weighted contextual API dependency graphs and graph similarity metrics, the small differences between two ZitMo variants designed to communicate via different channels were used as a case study on the system's effectiveness.

In the recent past a new threat emerged called Emotet, characterized by its rapid evolution and versatility together with its use of multi-layer obfuscation and anti-analysis techniques, that quickly became a serious concern [2], [3]. One of the first reports on Emotet was presented by [29], and the significant divergence of its credential harvesting method (i.e. network traffic sniffing) from prior banking Trojans was noted. The multi-layer deobfuscation process for its dropper was analyzed by [30], noting the complexity of its obfuscation and anti-analysis techniques, in addition to the common pitfall of creating deobfuscation scripts based on live samples that are subject to change. A more complete view of Emotet was presented by [31]; more details about its infection campaign were given, including the use of uninteresting topics on spam

emails, which helped gain their potential victims' trust, and the very short lifespan of compromised web servers hosting its payload. A number of reports on the evolution of Emotet were also presented by [4], [5], [32], in which it was shown that the bulk of C&C servers were located in the United States, while also noting the *malware-as-a-service* (MaaS) monetization scheme followed by Emotet's operators, and the utilization of email thread hijacking, in conjunction with spam emails, to propagate.

## III. ATTACK STRATEGIES

Banking Trojans proliferated during the past decade, as more and more financial institutions started to provide web access to their services [33], and threat actors employed them to intercept credentials or any sensitive information present on their systems. A number of attack vectors were exploited to spread such malware, through vulnerability exploitation and psychologically manipulative means (i.e. social engineering). Furthermore, the already significant threat posed by banking Trojans was magnified by their evolution to full crimeware services, as malicious actors seized the opportunity to utilize access to infected machines so as to mount further attacks by grouping them together in botnets, or for profit by offering their services to other malicious actors.

### A. Zeus

Zeus (also known as Zbot) offered the ability to attackers of varying skills and motivations to easily compile and deploy highly customized banking Trojans, while also allowing for the creation and management of a botnet [16]. The original variants of Zeus (i.e. versions 1 and 2) followed the *client-server* networking paradigm, with an architecture consisting of a central C&C server and a number of bots (i.e. infected hosts). It was distributed as a toolkit which included the C&C server files (written in PHP 5) and a bot executable builder for Windows systems. The full source code of version 2.0.8.9 was disclosed publicly in mid 2011 [34] which led to the creation of a number of variants, including Citadel, GameoverZeus, Ice IX, and KINS. Prior to its disclosure, the toolkit was available on underground forums; new versions were sold for a fee while older ones circulated for free.

The use of drive-by downloads triggered by exploitation frameworks as the primary infection vector was recorded as early as 2009 [15]. Their use involves the inclusion of malicious code on a website either through vulnerability exploitation (e.g. SQL injection, cross-site scripting) or through externally controlled elements (e.g. malicious advertisements). Apart from that, phishing schemes and other social engineering techniques were also used to spread Zeus.

Upon successful system infection, the installed bot could then perform a number of actions as instructed by the C&C, including: *hypertext markup language* (HTML) injection, transparent redirects, installation of certificates, interception of login credentials from *file transfer protocol* (FTP) and *post office protocol 3* (POP3) traffic, key logging, and screenshot capture. In order to capture credentials, HTML injection on a number of predefined bank websites was performed. Furthermore, the C&C offers the ability to receive status reports from the bots (*see* Figure 1), to update the bot configuration, and to manually execute a script of commands on a specific system. All communications happen over HTTP and all data is encrypted with RC4, using a unique key per botnet.

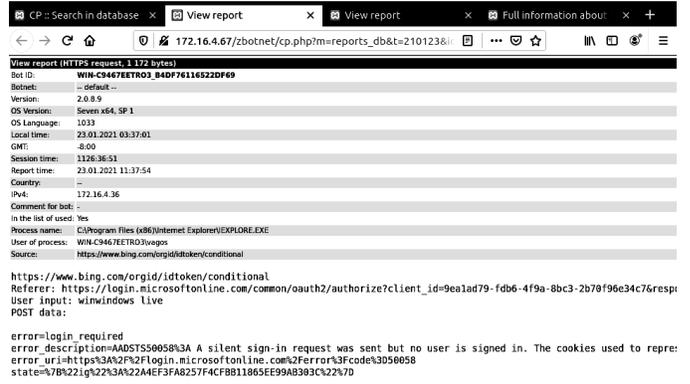

Fig. 1. Zeus bot report as viewed from the C&C control panel.

### B. ZitMo

Zeus-in-the-Mobile complemented the Zeus toolkit by targeting mobile devices to intercept mTANs and other sensitive information received by SMS messages. First detected in late 2010 [10], ZitMo originally targeted most of the popular early mobile platforms, including Symbian, Windows Mobile, BlackBerry and Android. A number of variants were discovered with their identifying characteristic being the targeted platforms and the communication vector. The first variant reported by [10], [24] was related with the SMS Monitor spyware [25] and communicated solely via SMS messages with a number of C&C phone numbers. This variant also had the ability to receive commands, for instance to receive additional numbers to forward messages to—with the exception of early Android variants which gained this ability a few months later. A second variant of ZitMo was able to communicate both via SMS messages and HTTP requests [26], [27]. This version was also adapted for use with other crimeware toolkits (e.g. URLZone), with one of its incarnations being known as ZertSecurity [35].

The installation of ZitMo was one of the steps of a Zeus attack. At some point when the victim browsed to a bank website, a message would appear after a successful login asking for the installation of a certificate installer (for the original variant and ZertSecurity) or of a security solution (as for example in Figure 2). ZitMo would then ask for the bank account number and would proceed to provide a passcode to be entered in the Zeus-modified bank website. This passcode would then be used to correlate information from the newly installed ZitMo instance with the specific bot and exploit attempt. Another infection vector involved spam emails [36] supposedly originating from a bank, asking for the installation of ZertSecurity alongside its existing mobile banking application, while also providing detailed instructions on how to install third-party applications.

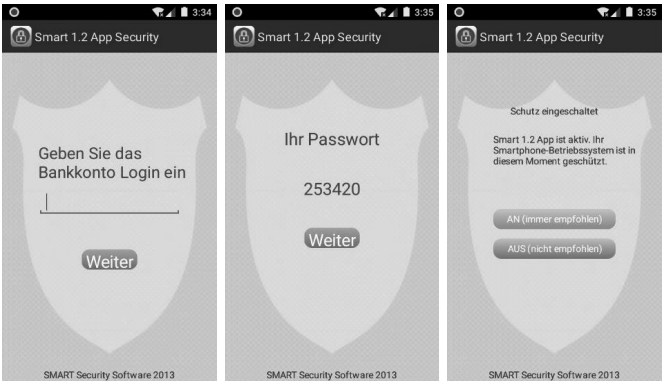

Fig. 2. The three views of the second Android ZitMo variant.

*C. Emotet*

Emotet originally started as a banking Trojan (then known as Geodo) with the ability to transparently sniff network traffic to intercept credentials sent to banking websites [29]. A few years later, it was converted from a banking Trojan to a modular platform capable of installing other malicious payloads, thus switching to a *malware-as-a-service* (MaaS) monetization scheme. It also had the ability to form organize its infected machines into botnets; up to the point of its infrastructure takedown, three separate C&C servers (with separate distribution campaigns, encryption keys, and so on) were operated, referred to as Epoch 1, Epoch 2, and Epoch 3. The complex structure, its rapid and near-constant evolution, along with the employment of multiple layers of obfuscation and other anti-analysis measures made its detection from antimalware solutions quite difficult, as information extracted from captured samples could be rendered obsolete in a short period of time [30].

Customized batches of spam emails were used to spread the dropper, changing the content slightly to add randomness so as to avoid signature matching. The subject choice for these messages was characterized by [31] as "the most mundane topics imaginable", hoping to increase the probability of a victim opening the document or following a link to download it. Emails were also sent to correspondents of the victim as responses to existing email threads (i.e. thread hijacking), to gain more credibility and propagate without raising suspicion. The dropper is in the form of either a Microsoft Office document with embedded macros, or sometimes downloaded from a link contained in a PDF file. Their content prompted the potential victim to disable security features against macro execution, by claiming technical problems displaying its contents. Emotet binaries were hosted on compromised web servers which were taken down within hours of their discovery [31].

After the malicious document was opened and the system was successfully infected, the Emotet binary proceeded to collect information about the system and its user, while it also downloaded and installed the second-stage payload Information collection was performed through plugins, some of which were existing freeware tools (e.g. NirSoft tools). Their functionality included, among others, the ability to extract email contacts from Outlook, and to scan for writable SMB file shares to propagate. The C&C server was contacted frequently for updates, further commands, or to send collected data. Most commonly, communications happened over HTTP requests by passing AES-encrypted data in the cookie field.

## IV. TECHNICAL DETAILS

In the years after the abandonment of Zeus and the initial detection of Emotet, measures against malware were taken by computer manufacturers while antimalware solutions matured significantly. A more detailed description of the actions performed during the initial execution of each malware can provide a better understanding of the change in tactics employed by malware to successfully spread and evade detection.

*A. Zeus*

Few differences were noted between Zeus version 2.0.6.5, presented by [16], and version 2.0.8.9, used to test the IRS in Section V. A bot executable assembled by version 2.0.8.9 of Zeus, upon its initial execution (by the victim, or by an exploited vulnerability), will proceed to unpack itself and create a new copy to a randomly-named folder under `%AppData%`. System information will be collected and command line arguments (`-f`, `-i`, `-n`, or `-v`) will be processed, while a 512-byte block from the `.data` segment will be decrypted using RC4—this block contains the flag differentiating the original bot executable from its copy. A unique mutex for the specific bot executable will be created, to avoid reinfection from the same botnet, while allowing for multiple botnets to infect the same user. The dropping routine of the original bot executable will be deobfuscated, using XOR operations and a 4-byte key from the aforementioned block, in addition, a new registry key with a random name will be created at `HKCU\Software\Microsoft` to store the bot configuration. In preparation for the final stage, a new 512-byte block will be assembled, storing information about the system, the specific infection, and the RC4 key used when communicating with the C&C—as defined by the botnet operator during the C&C server installation procedure. Finally, this block will be written on the `.data` segment of the copy, which is then named with a random name and executed.

This newly created file, which is set to be executed every time the system is booted, will delete the original bot executable, and will then proceed to determine whether it has been tampered, by comparing its decrypted 512-byte block with the results taken by repeating the block assembly routine. After the checks are done, the executable will inject its code to every process it has the rights to modify, hook a number of their API calls, and terminate. A number of system processes (e.g. `taskhost.exe` or `explorer.exe`) will also host a number of extra threads, for network communications, to download the configuration from the C&C, and to monitor the bot's registry keys.

The network communication patterns of Zeus (under normal operation) are relatively simple, as seen in Figure 3, with the bot communicating once with the C&C to receive the configuration file, and then performing periodic pings to the C&C for status updates. This configuration file, easily customizable by the botnet operator, includes links to critical files on the C&C server, and required information to perform HTML injection on the specified websites.

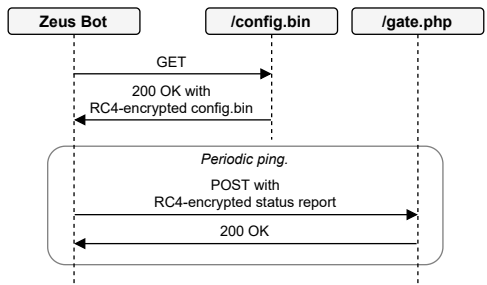

Fig. 3. Communication patterns of the Zeus ver. 2.0.8.9.

All exchanged data, including the configuration file itself, is obfuscated using XOR operations and then encrypted with RC4 [17]. A comment on this choice, made in the disclosed instruction manual[1], makes clear that the sole purpose of encryption is to avoid alerting network IDS systems, not data security nor protection against false information injection attacks. Furthermore, to avoid detection from signature-based antimalware solutions the bot executables of version 2 used randomization, both when creating the final bot executable (changing its hash value) and during operation. It was also common practice among botnet operators to pack the bot executables using a packer of their choice, in addition with the existing Zeus obfuscation processes [16].

*B. ZitMo*

The second ZitMo variant presented here was analyzed by the authors and a Python C&C server was written to understand its communication patterns, as existing works on it and ZertSecurity do not provide many details. As demonstrated from the message flow presented in Figure 4 (whose message contents are listed in Table I), the second Android ZitMo variant[2] exhibits a relatively simple behavior. Only two requirements need to be fulfilled for the infection to be successful: all requested permissions[3] must be granted, and the victim must complete steps #1–4.

Starting from the initial execution of ZitMo by the victim, the C&C receives a message with an empty `login` field, the victim's phone number, the device's *international mobile equipment identity* (IMEI) number (in the `devid` field), and the CRC32 value of the hardcoded C&C URLs (in the `dd` field). The server then, either responds with a new list of URLs (message #2), or with a message indicating that no changes are necessary (message #6). The victim is prompted to input their bank account number which is reported back to the server (message #3), and is followed by an empty response (message #4). After this point, ZitMo will proceed to periodically ping the C&C for an updated list of URLs (message #5), on which the server will respond either negatively (message #7), or positively (message #2). The ping interval is set to 15 minutes.

---
[1] A detailed manual for version 2.1.0.0 (last updated: Mar. 20, 2011) was posted in: `pastehtml.com/view/1ego60e.html`
[2] The sample analyzed in this section was used on a URLZone campaign; no significant difference with other ZitMo samples (or ZertSecurity) was noted.
[3] These are: `RECEIVE_BOOT_COMPLETED`, `READ_PHONE_STATE`, `ACCESS_NETWORK_STATE`, `INTERNET`, `RECEIVE_SMS`, `SEND_SMS`, and `WAKE_LOCK`.

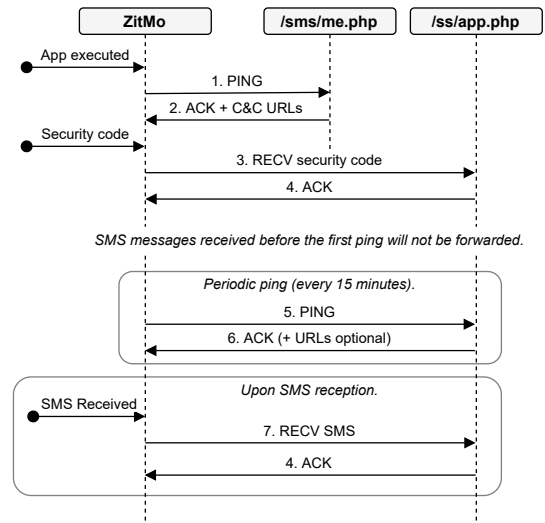

Fig. 4. Communication patterns of the second Android ZitMo variant.

All messages are encrypted with AES in ECB mode using a pre-shared key (`0523850789a8cfed`), common for most variants of ZitMo or ZertSecurity, and the message is padded with space characters. The encrypted message is finally encoded in Base64 and transmitted over HTTP. All valid C&C responses (for both HTTP and SMS communications) are terminated with `&Sign28tepXXX`, which as seen in message #4 is also the empty response. Settings are also stored in a text file at `/data/data/com.guard.smart/cfg.txt`, to be reloaded in case the device is rebooted.

*C. Emotet*

The latest version of Emotet, as described in [5], [32], has many differences when compared to older versions, but its basic behavior has not changed significantly—the reader is also encouraged to contrast the reports presented in Section II. After the victim has downloaded the malicious Microsoft Office document, and with macro execution enabled (as prompted by the document itself), the attached *Visual Basic for Applications* (VBA) macro script will proceed to deobfuscate the second stage of the dropper, a PowerShell script which will then download the Emotet executable. Both scripts are obfuscated: the VBA script primarily uses dead code insertion and string construction techniques, to confuse inexperienced malware analysts and sandboxes, while the PowerShell script is assembled via string pattern replacement. As reported by [5], the final deobfuscated VBA script is just 11 lines, from over 400 lines originally.

TABLE I
MESSAGES EXCHANGED BETWEEN ZITMO AND THE C&C.

| # | Message |
|---|---|
| 1 | `services=timer&login=&phone=+15555215554&devid=358240051111110&dd=C80B059D&` |
| 2 | `0&http://172.17.0.1:8000/ss/app.php&Sign28tepXXX` |
| 3 | `services=login&login=123456789&phone=+15555215554&devid=358240051111110&` |
| 4 | `&Sign28tepXXX` |
| 5 | `services=timer&login=123456789&phone=+15555215554&devid=358240051111110&dd=ADFC7D64&` |
| 6 | `1&Sign28tepXXX` |
| 7 | `services=sms&text=OTP+for+transaction+%235356323274+is+163572.&number=3085550174&login=123456789&` |

The downloaded Emotet executable contains the encrypted next stage in its .rsrc section, which in turn contains the loading routine (as the pointer to the newly-decrypted contents will be called as a function pointer) and the executable of the next stage. The decrypted executable will run through the loaded DLLs of its parent process to load its required functions, by matching them to a list of custom-made hashes (produced from the library and function names). Afterwards, this decrypted executable will proceed to unpack the final Emotet payload.

The final payload exhibits different behavior depending on its file timestamp. If it determines that it runs for the first time, either because it is dropped during the initial infection, or because it has been updated, it will proceed to clear any existing registry keys, move itself to the desired path (as specified by its input arguments, or to a random one otherwise), and will modify its timestamp to an older date. Differently, it will proceed to create a thread to monitor its executable filename, and its communications with the C&C will begin. Upon successful contact, a new autorun registry key will be created and the desired payloads will be executed. This executable uses a number of obfuscation techniques as well, most notably string obfuscation using XOR operations and a 4-byte key, and control flow graph flattening—as its code is structured in if-else blocks contained inside a large while loop (i.e. the dispatcher), whose context is a variable updated at the end of each internal block. C&C communications, on most Emotet versions, use a two-layer encryption scheme, in which the entire message (payload and headers) is encrypted with a hadcoded RSA public key (different per Epoch), while its payload contents are encrypted with an AES-128 key generated by the Emotet executable and shared with the C&C during initial contact.

## V. Experimental results

To combat these malware threats, in addition to others targeting different attack vectors and systems, a multi-level approach to network security is required, to combine both host-based and network-based solutions. For the latter, a number of approaches to network security modeling and attack mitigation have been proposed. These approaches, implemented by IRS solutions, aim to mitigate unknown or sophisticated attacks targeting their protected network. By modeling and assessing the network security state, which is updated by alerts produced by network IDS solutions (both signature-based and AI-based), IRS solutions are able to choose the most optimal and effective mitigation actions to respond to cyber attacks.

### A. Intelligent intrusion response system (iIRS)

To assess the success of IRS solutions against malware threats, the *intelligent intrusion detection system* (iIRS) presented in [37], [38] has been tested on a network with a Zeus-infected Windows host present. This system is based upon a graph-based GNSM of the network, the *Bayesian attack graph* (BAG) form, which allows for the application of probabilistic cyber attack models.

Information retrieved from the network router and Nmap scans (with the Vulners NSE plugin[4]) is used to generate Datalog tuples for the *Multi-host, Multi-stage Vulnerability Analysis Language* (MulVAL) reasoning engine [39], whose output is then mapped to the underlying BAG model. The MulVAL-produced *logical attack graph* (LAG) is comprised by three types of nodes: OR & LEAF nodes representing possible states of the network devices (e.g. access relations, or vulnerabilities), that is, capabilities an attacker may acquire; and AND nodes representing conjunctive relations between OR & LEAF nodes. Edges represent directed transitions from a number of precondition states to a number of post-condition states. When mapped to a BAG, nodes can be seen as random variables connected in pairs via conditional dependencies.

Furthermore, to defend against network attacks, the game theoretic approach presented in [40], based on a discrete-time *partially observable Markov decision process* (POMDP), is used to calculate the optimal mitigation action based on the system's belief about the network state. This belief is expressed by assigning a probability of exploitation to all OR & LEAF nodes of the BAG (i.e. the security conditions), and is updated by matching IDS alerts to their respective AND nodes (i.e. exploits). Three matching levels were implemented and used in cascading order: the most specific matched an IP address, a hostname, a network port and protocol; the next one matched everything except for the port; whereas the least specific one only matched an IP address and hostname.

The POMDP model implemented by the iIRS considers an attacker aiming to exploit system vulnerabilities to reach a goal state, and a defender attempting to prevent the attacker's progression. The only action a defender can take, in this version of the iIRS, is to deploy `iptables` firewall rules either blocking a specific connection between two network hosts (specific rules), or all connections originating from a specific host (general rules) [38].

### B. Simulation setup and attack scenario

The scenario under which the iIRS was tested involved the infection of a Windows host with a Zeus ver. 2.0.8.9 bot, and its successful communication with a C&C server positioned outside of the local network—Figure 1 was captured during the execution of this scenario. Two variations of this scenario were performed: one starting from the initial infection step, at the moment the targeted Windows machine downloads the bot executable; and one starting from the introduction of an already infected Windows machine to the network.

---

[4]github.com/vulnersCom/nmap-vulners

Testing was performed on a virtualized network consisting of two sub-networks (one internal, one external) and one gateway router. For this scenario only three out of the eight machines were involved:

- The gateway router, which hosts the network discovery tools, is responsible to enforce the firewall rules, and hosts a Suricata IDS instance—located at 192.168.0.1 internally and 172.16.4.36 externally.
- The Windows 7 SP1 host—located at 192.168.0.17.
- The Zeus C&C Ubuntu host—located at 172.16.4.67.

In this configuration, the iIRS was hosted on a machine located inside the gateway's local network, at an address range ignored by the network discovery tools and the IDS. The gateway's network discovery tools provided all necessary network topology information for the generation of the GNSM, while the IDS instance generated alerts based on known malicious traffic pattern matching.

### C. Detection and mitigation

A number of alerts were expected to be received during testing, for both variations, as signatures exist for both the bot executable and its traffic patterns (as presented in Figure 3); in addition with the existence of more general signatures which may also apply.

During the first scenario variation, starting from the initial infection, seven alerts were generated immediately when the bot executable was transferred through the gateway—five were signature matches[5] (2011967, 2016141, 2019714, 2018959, and 2021076), while two were informative about the file and its related HTTP communications. During the second scenario variation, with the machine already successfully infected, During the second variation, with the bot already attempting to communicate with the C&C, seventeen alerts were generated upon the first communication attempt—eleven were signature matches (2016858, 2017930, 2019141, 2022986, and 2018358 twice; 2016173 once), while six were informative about the `gate.php` requests.

In both variations, in response to the received alerts the following general rules were selected:

```
iptables -A INPUT -s 192.168.0.17 -j DROP
iptables -A OUTPUT -s 192.168.0.17 -j DROP
```

These rules, blocking all communications originating from the infected Windows machine, were chosen as it was not possible for the resulting GNSM to model the ports opened after its generation. These were, the port opened by the victim's browser to download the bot executable (for the first variation), and the constantly changing ports opened by the bot to communicate with the C&C (for the second variation).

---

[5]On the Emerging Threats rule set: doc.emergingthreats.net

### D. Future work

As demonstrated from the previous sections, with the exception of the single-purpose ZitMo, both Zeus and Emotet exhibit a number of common behavioral patterns:

- Appropriately complex obfuscation techniques are used against automated analysis tools and to cause significant delays to reverse engineering efforts.
- Randomization is used to bypass signature-based antimalware solutions installed on infected systems, and to cause uncertainty to network IDS solutions.
- Communication with their respective C&Cs happens through common networking protocols (e.g. HTTP), using encryption to avoid detection by signature-based network IDS solutions.

The last two points are most important for network-based IRS solutions. Randomization needs to be taken into account, as it may cause advanced IDS systems to produce noisy alerts (i.e. with an unacceptable number of false alerts), thus rendering their usage difficult or even impossible. Network traffic obfuscation techniques, if used properly, can hinder the work of IRS solutions as well, by rendering existing signatures unusable and making the creation of new ones significantly harder with every new malware variant.

In future works, mobile devices, targeted by malware like ZitMo, and more complex malware like Emotet, should be tested with the existing iIRS implementation, to discover areas on which its models are lacking. Improvements to the GNSM generation process need to be made in order to reflect the state of the network in real time, or at least to reflect the changes made between two discrete time steps of the POMDP—as indicated by the inability of the iIRS to target the specific port on which the bot communicates. Moreover, integration with more advanced network IDS solutions, for instance with anomaly detection or AI-based systems, must be considered to ensure that the iIRS is able to deal with novel and rapidly evolving threats—as seen with the very short life of Emotet's hosting servers [31], and its multitude of payload configurations.

## VI. CONCLUSION

To track the evolution of banking Trojans during the last decade, two representative examples from the beginning and the end of the past decade were presented. Both Zeus (along with its mobile companion ZitMo) and Emotet were developed with the same aim, to intercept credentials and sensitive information related to financial transactions, but evolved in entirely different directions. After the disclosure of its source code Zeus spawned a number of more successful variants and was ultimately abandoned, having served its purpose. On the other hand, Emotet evolved rapidly and pivoted to monetizing its access to infected machines to other malicious actors.

Three common behavioral patterns were identified: 1) use of obfuscation techniques to bypass automated analysis tools and delay manual analysis efforts, 2) use of randomization on various places to bypass signature-based antimalware solutions, and 3) use of common networking protocols to communicate through encrypted messages with their C&C server. The last two patterns are recognized as important enough for the implementation of an effective *intrusion response system*.

The effectiveness of the network-based *intelligent intrusion response system* (iIRS), presented in [37], was tested on a virtualized network with a Windows machine targeted by Zeus (ver. 2.0.8.9). This system processes information retrieved from the gateway router and Nmap scans to generate the necessary input for the MulVAL reasoning engine, whose outputted graphical model is then mapped to a *Bayesian attack graph*, which becomes the basis upon which a discrete-time POMDP, a game theoretic model of cyber attacks, will be run. This process receives network alerts from the Suricata IDS instance located on the gateway, to update its belief about the security status of the network. This process conceptualizes the existence of two actors, an *attacker* aiming to reach a desired goal state (i.e. a specific graph node), and a *defender* trying to prevent the attacker's progression by employing `iptables` firewall rules to change the interconnectivity of the hosts.

Experimental results for the aforementioned scenario were encouraging, as the iIRS managed to successfully isolate the Windows machine both upon the detection of the Zeus bot executable being downloaded, and upon the detection of bot to C&C communications. Although, future works need to further test the effectiveness of IRS solutions against novel threats, and optimize a number of modeling and algorithmic aspects which hurt their effectiveness.